# Incremental Value and Interpretability of Radiomics Features of Both Lung and Epicardial Adipose Tissue for Detecting the Severity of COVID-19 Infection

Ni Yao, M.S.[1*], Yanhui Tian, B.S.[1*], Daniel Gama das Neves, MD[2,3], Chen Zhao, M.S.[4], Claudio Tinoco Mesquita, M.D, Ph.D.[2], Wolney de Andrade Martins, MD, Ph.D.[3,5], Alair Augusto Sarmet Moreira Damas dos Santos, MD, Ph.D.[2,3], Yanting Li, Ph.D.[1], Chuang Han, Ph.D.[1], Fubao Zhu, Ph.D.[1], Neng Dai, M.D., Ph.D.[6,7], Weihua Zhou, Ph.D.[4,8]

[*]These authors contributed equally to this work and should be considered co-first authors.

[1]School of Computer and Communication Engineering, Zhengzhou University of Light Industry, Zhengzhou, Henan, China
[2]Department of Radiology, Universidade Federal Fluminense, Rio de Janeiro State, Brazil.
[3]DASA Complexo Hospitalar de Niterói, Rio de Janeiro State, Brazil
[4]Department of Applied Computing, Michigan Technological University, Houghton, MI, USA
[5]Department of Cardiology, Universidade Federal Fluminense, Rio de Janeiro State, Brazil.
[6]Department of Cardiology, Zhongshan Hospital, Fudan University, Shanghai Institute of Cardiovascular Diseases, Shanghai, China
[7]National Clinical Research Center for Interventional Medicine, Shanghai, China
[8]Center for Biocomputing and Digital Health, Institute of Computing and Cybersystems, and Health Research Institute, Michigan Technological University, Houghton, MI, USA.

Corresponding authors:
Weihua Zhou, PhD
Department of Applied Computing, Michigan Technological University 1400 Townsend Dr, Houghton, MI 49931, USA.
E-Mail: whzhou@mtu.edu

Neng Dai, MD, PhD
Department of Cardiology, Zhongshan Hospital, Fudan University, Shanghai Institute of Cardiovascular Diseases, National Clinical Research Center for Interventional Medicine, 180 Fenglin Road, Xuhui District, Shanghai 200032, China.
E-Mail: niceday1987@hotmail.com

**Abbreviated title: Detecting COVID-19 Severity with Lung and Epicardial Adipose Tissue Radiomics Features**

# ABSTRACT

Epicardial adipose tissue (EAT) is known for its pro-inflammatory properties and association with Coronavirus Disease 2019 (COVID-19) severity. However, current EAT segmentation methods do not consider positional information. Additionally, the detection of COVID-19 severity lacks consideration for EAT radiomics features, which limits interpretability. This study investigates the use of radiomics features from EAT and lungs to detect the severity of COVID-19 infections. A retrospective analysis of 515 patients with COVID-19 (Cohort1: 415, Cohort2: 100) was conducted using a proposed three-stage deep learning approach for EAT extraction. Lung segmentation was achieved using a published method. A hybrid model for detecting the severity of COVID-19 was built in a derivation cohort, and its performance and uncertainty were evaluated in internal (125, Cohort1) and external (100, Cohort2) validation cohorts. For EAT extraction, the Dice similarity coefficients (DSC) of the two centers were 0.972 ($\pm$0.011) and 0.968 ($\pm$0.005), respectively. For severity detection, the hybrid model with radiomics features of both lungs and EAT showed improvements in AUC, net reclassification improvement (NRI), and integrated discrimination improvement (IDI) compared to the model with only lung radiomics features. The hybrid model exhibited an increase of 0.1 ($p<0.001$), 19.3%, and 18.0% respectively, in the internal validation cohort and an increase of 0.09 ($p<0.001$), 18.0%, and 18.0%, respectively, in the external validation cohort while outperforming existing detection methods. Uncertainty quantification and radiomics features analysis confirmed the interpretability of case prediction after inclusion of EAT features.

# Incremental Value and Interpretability of Radiomics Features of Both Lung and Epicardial Adipose Tissue for Detecting the Severity of COVID-19 Infection

## 1. Introduction

The COVID-19 pandemic, caused by severe acute respiratory syndrome-coronavirus-2, has resulted in a significant global health crisis [1, 2]. While primary studies have primarily focused on lung complications and radiomics features for detecting the severity of COVID-19, there is a need to consider additional factors such as inflammation and the role of epicardial adipose tissue (EAT) in the progression of the disease [3-5].

Inflammation is intricately linked to the progression of COVID-19 and can provide more value for detecting severity based on lung features [5, 6]. EAT, located between the myocardium and pericardium, is a recognized source of pro-inflammatory mediators [7, 8]. The inflammatory response plays a crucial role in the development and progression of COVID-19 [9]. A correlation between EAT volume and the severity of COVID-19 has been established through various studies [10-15], thereby aiding in the assessment of the risk of disease advancement. Existing studies have primarily focused on the volume of EAT, but with the advancement of radiomics, exploring additional radiomics features of EAT can provide valuable insights [16, 17]. The study on the impact of EAT on the severity detection of COVID-19 also helps to provide new ideas for the study of common diseases.

The traditional method of EAT extraction demands a laborious and time-consuming annotation process. Simultaneously, existing methods [18, 19] lack consideration for EAT positional information, therefore there is a need to enhance the overall performance of artificial intelligence methods to ensure high segmentation accuracy in automatic extraction of EAT.

In addition, in the context of medical diagnostics, uncertainty plays a crucial role in enhancing the interpretability of predictive models[20, 21]. Quantifying and communicating uncertainty in medical diagnostics are crucial for enhancing the interpretability of predictive models and gaining the trust of healthcare providers and patients.

To achieve these objectives, the study proposes a novel three-stage method for the automatic extraction of EAT, explores the incremental value of radiomics features of both lung and EAT in detecting the severity of COVID-19 infection, and verifies the interpretability of the model based on uncertainty and radiomics. This research has the potential to provide valuable insights into the role of EAT in COVID-19 severity and enhance the overall understanding of the disease.

## 2. Materials and Methods

*2.1 Study population*

The retrospective study included patients from two centers between January 2020 and July 2020 (**Fig. 1**). There were 415 consecutive patients (371 mild cases and 44 severe cases) with confirmed COVID-19 cases in Shanghai Public Health Center (Cohort 1) [5] and 100 consecutive patients (50 mild and 50 severe) with confirmed COVID-19 cases in Brazil Niteroi Hospital (Cohort 2).

For EAT extraction study, the heart contour of 47 subjects in cohort 1 (27 mild and 20 severe) and 15 subjects in cohort 2 (8 mild and 7 severe) was manually drawn by experienced operators on LabelMe (Version 4.5.9) and used to train the segmentation model.

For the classification modeling phase of mild and severe cases, the EAT of the remaining 368

subjects from cohort 1 and 85 subjects from cohort 2 was extracted automatically using a trained deep learning model. 415 patients (371 mild cases and 44 severe cases) with confirmed COVID-19 from cohort 1 were randomly divided into a derivation cohort (n=290, 260 mild cases, 30 severe cases) and an internal validation cohort (n=125, 111 mild cases, 14 cases) (Fig. 1). Each case contained its clinical information and a set of chest computed tomography (CT) images in cohort 1. 100 patients (50 mild and 50 severe) from cohort 2 were deemed as an external validation cohort (Fig. 1). Each case contains a set of chest CT images in cohort 2.

This study was approved by the Ethics Committee of the Shanghai Public Health Clinical Center and the Brazil Public Health Clinical Center Ethics Committee.

*2.2 Clinical data collection*

For mild-severe classification criteria, patients admitted to the ICU and infirmary were diagnosed with severe COVID-19, and patients who only required treatment in the emergency department were diagnosed with mild COVID-19.

*2.3 Chest CT procedures*

CT scans were performed using Toshiba Aquilion 64 CT scanner in cohort 1. Image acquisition parameters consisted of 120 kV and 114.1 mAs. The thickness of the CT scanned slices was 5mm, and the in-plane pixel size ranged from 0.579mm to 0.935mm. The number of chest CT slices for each patient ranged from 52 to 70. CT scans were performed using a Siemens Somatom Sensation 64 CT scanner in cohort 2. Image acquisition parameters consisted of 120 kV and 94 mAs. The thickness of the CT scanned slices was 1mm, and the in-plane pixel size ranged from 0.669mm to 0.815mm. The number of chest CT slices for each patient ranged from 300 to 370.

Due to the difference in CT data acquisition imaging between the two cohorts, the segmentation method was used to train the model in the two cohorts.

*2.4 Lung Segmentation*

The lung segmentation was achieved according to the research method of Zhao et al. [22], and the segmentation results of lung were confirmed by radiologists. Unsatisfactory results will be manually adjusted.

*2.5 Automatic Extraction of EAT*

The method for the automatic extraction of EAT is structured into three main modules, each serving a specific function in the segmentation process. The process is illustrated in Fig. 2. Firstly, object detection is used as a guiding module in the initial stage to obtain positional information, which contributes to improved accuracy in subsequent segmentation. Following this, the results from object detection are then leveraged to conduct binary segmentation of the heart, distinguishing between the myocardium and endocardium, as well as the background. This step is carried out within the segmentation network, ultimately producing the contour of the heart. Finally, the third stage involves the implementation of thresholding and smoothing techniques within the heart contour to isolate and extract the EAT.

*2.5.1 Object Detection*

In order to address the time-consuming manual screening of each patient's heart start and end frames,

object detection is incorporated as a guide module to enhance the efficiency of heart segmentation. The YOLO-V5 network [23] is utilized for object detection, with the training samples consisting of images containing the heart. The input image size for the YOLO-V5 network is set to 512×512×1, and the corresponding labels include the normalized coordinates of the heart centroid (horizontal and vertical) within the image, as well as the width and height of the heart region. By passing the image through this bootstrap module, the network can output information regarding the heart's location. If the image contains the heart region, it will be identified as such; otherwise, it will be classified as not containing the heart region. This step enables automatic screening of images to identify those that contain the heart region for each patient. Subsequently, these identified images are used as input for the subsequent steps in the segmentation process.

*2.5.2 Heart Segmentation*

In order to achieve more accurate determination of the epicardial adipose tissue (EAT) range, a deep learning network was employed to automatically segment the heart contour in each patient image, before proceeding with the extraction of EAT. The heart contour segmentation model utilized in this study is based on the well-established U-Net network [24]. The U-Net network is a two-dimensional model with an input image size of 512×512×1. Upon processing an input image, it produces an output in the form of a probabilistic map that represents the heart contour. In this map, a value of 1 indicates the presence of the heart contour and its internal regions, while 0 represents the background. This approach ensures the precise delineation of the heart structure, thereby facilitating the subsequent extraction of EAT.

*2.5.3 Threshold Extraction*

As used in the generation of the reference, the threshold for adipose tissue has been defined as a Hounsfield Unit (HU) of -190HU to -30HU. EAT was obtained by threshold extraction and median filtering in the heart contour results obtained from the segmentation network.

***2.6 Severity Detection***

*2.6.1 CT radiomics features extraction*

In cohort 1 and cohort 2, Each patient's lung and EAT regions were formed into three-dimensional images, and 120 radiomics features [25] of both lung and EAT were extracted, respectively.

*2.6.2 Feature Selection*

In a derivation cohort of 290 patients with COVID-19, univariate analysis and Pearson correlation analysis were used to identify predictors of COVID-19 severity classification from 120 lung and 120 EAT radiomics features. This analysis process refers to the method of Zhao et al. [22]. Univariate logistic regression analysis was used to select features with significant differences. Then Pearson correlation analysis between features was conducted and strong correlation features were excluded with the selection threshold combined with the final number of included features.

*2.6.3 Classification Model Modeling*

In the derivation cohort (n=290), a hybrid model based on seven basic machine learning models was built. The basic machine learning models include Logistic Regression (LR)[26], Support Vector Machine (SVM)[27], Random Forest (RF)[28], Adaptive Boosting (AdaBoost)[29], Extreme Gradient

Boosting (XgBoost)[30], Light Gradient Boosting Machine (LGBM)[31], and Gradient Boosting Decision Tree (GBDT)[32]. The hybrid model utilizes the mean value of the predictions of the basic models as its prediction result, and the standard deviation of the predictions of the basic models is used as the uncertainty quantization of the prediction result.

The performance improvement of the COVID-19 severity classification model was subsequently validated using lung and EAT radiomics features in both an internal validation cohort (n=125) and an external validation cohort (n=100). The interpretability of the model was analyzed based on the radiomics features and the uncertainty quantization.

The uncertainty of prediction results was quantified into six levels ranging from 0 to 0.1, 0.1 to 0.2, 0.2 to 0.3, 0.3 to 0.4, 0.4 to 0.5, and 0.5 to 1. A lower uncertainty quantification value closer to 0 indicated a higher confidence in the model's prediction for a case, while a value closer to 1 represented higher uncertainty and less confidence in the model's prediction for a case. The uncertainty quantification results of the hybrid model combining radiomic features of both lung and EAT were evaluated in terms of their quantification levels in the internal validation cohort and external validation cohort, respectively.

Additionally, the model was verified by comparing its performance with state-of-the-art methods.

*2.7 Evaluation metrics*

Precision and recall were used to evaluate the accuracy of the object detection model. DSC and Hausdorff distance (HD) were used to assess the validity of the segmentation model. The ROC curve analysis was used to calculate the AUC and select a cut-off value of severity classification, and calculate the accuracy, sensitivity, and specificity of detecting severity. The estimated 95% Confidence Interval (CI) was calculated using 1000 bootstrap estimates. The improvement degree of the prediction effect of the model was evaluated according to the NRI and IDI. Feature extraction, analysis, modeling, and evaluation were performed using Python (Version 3.7) and Pyradiomics (Version 3.0.1).

# 3. Results

*3.1 Patient characteristics*

In the derivation cohort, 30 (10.34%) patients were diagnosed with severe COVID-19. The mean age of mild and severe cases was 39.98±15.39 and 59.23±14.86, respectively, and 48% (n=260) and 53% (n=30) were male, respectively. The baseline characteristics of patients in the derivation cohort are displayed in Table 1. There were no significant differences (p>0.05) in gender, white blood cell count, potassium, and lactic acid.

In the internal validation cohort, 14 (11.2%) patients were diagnosed with severe COVID-19. The mean age of mild and severe cases was 40.46±15.23 and 59.71±14.46, respectively, and 59% (n=111) and 79% (n=14) were male, respectively (Table 1). There were no significant differences (p>0.05) in gender, white blood cell count, lactic acid, and creatinine.

Fifty patients (50%) were diagnosed with severe COVID-19 in the external validation cohort. The mean age of mild and severe cases was 46.22±7.38 and 62.38±16.14, respectively, and 56% (n=28) and 60% (n=30) were male, respectively. There were no significant differences (p>0.05) in gender.

*3.2 EAT Extraction*

The EAT extraction process utilized various methods to train segmentation models and compared the performance using Dice Similarity Coefficient (DSC) and Hausdorff Distance (HD) in two test sets.

The results, as presented in Table 2, demonstrate that the segmentation method employing YOLO-V5+U-Net outperforms existing methods significantly. The U-Net network predicted all images as background, with a resulting DSC of 0.629. This led to all prediction results being blank negative samples, making it impossible to calculate HD. Visual comparison between EAT segmentation results annotated by experts and those predicted by the model is illustrated in Fig. 3. Furthermore, a comparative test was conducted with a three-dimensional V-Net [33] network to explore whether object detection combined with a two-dimensional network would enhance segmentation. The experimental results revealed that while the 3D network can avoid the need to select CT images by object detection, its accuracy in segmenting image details is not as high as that achieved after object detection using the 2D network. As a result, the 2D network was chosen over the 3D network.

The proposed method demonstrated superiority over existing methods in two cohorts, attributed to the influence of negative samples provided by the object detection module prior to segmentation, which improved segmentation accuracy. Notably, the operational time for segmenting and extracting EAT from a patient's chest CT data on a standard computer was less than 20 seconds, a significant reduction compared to the approximately 20 minutes required for manual extraction by a specialist. This efficiency improvement represents a substantial time-saving advantage, facilitating more convenient EAT extraction and meaningful research endeavors.

*3.3 Feature selection and diagnostic model*

One hundred and twenty radiomics features were extracted from lung and EAT, respectively. Feature selection is performed in the derivation cohort. Seventy-five lung and forty-two EAT radiomics features were significantly different in severity classification. The order of feature selection was ranked from high to low according to the AUC score of the univariate analysis. The feature importance scores are shown in Fig. 4. The features with strong correlation were eliminated by correlation analysis among features (the correlation coefficient threshold was 0.75 in this paper). Finally, the remaining six lung radiomics features and four EAT radiomics features were included to build the diagnostic model (Table 3).

*3.4 Model performance in the validation cohort*

Table 4 and Fig. 5 shows the model performance for mild and severe classification in the internal and external validation cohort. As in previous studies, EAT was found to be related to the severity of COVID-19 in terms of severity classification. In this study, the incremental value of radiomics features of EAT integrated with lung for detecting COVID-19 severity is found and demonstrated in internal and external validation cohort. In the internal validation cohort compared with the hybrid model with only lung features, the hybrid model with lung and EAT radiomics features demonstrated better predictive efficacy; its NRI increased by 19.3% (p<0.001), and IDI increased by 18.0%(p<0.001). In the external validation cohort, compared with the hybrid model with only lung features, the hybrid model with radiomics features of both lung and EAT demonstrated better predictive performance; its NRI increased by 18.0% (p<0.001), and IDI increased by 18.0% (p<0.001).

What is more noteworthy is that regardless of the model, the performance of the model combining lung and EAT radiomics features was superior to the model with only lung radiomics features, which further validated the incremental value of EAT for COVID-19 severity detection.

Furthermore, a comparison was made with existing methods for COVID-19 severity diagnosis, as shown in Table 5. These studies employed different metrics to establish diagnostic models, including clinical features, radiological features of the lungs, and CT quantitative scores. Our method achieved optimal performance by incorporating radiomics features of both the lung and EAT.

*3.5 Interpretability analysis*

*3.5.1 Based on radiomics features*

The feature selection identified ZoneEntropy and Skewness as essential features for characterizing EAT. These features reflect the uncertainty or randomness in the size and gray distribution of the measurement area and the asymmetry of the value distribution. The adipose tissue attenuation value in EAT has a dynamic range [-190, -30], and these features highlight the instability of the adipose tissue attenuation value in the EAT region of the image. Fig.6 compared the EAT attenuation index between mild and severe COVID-19 patients and found that it was more unstable and closer to -30 HU in severe patients. This observation provides a more intuitive reflection of the association between EAT and COVID-19 severity. Previous studies [34] have shown that adipose tissue can influence inflammation, and this study hypothesized that COVID-19 affects EAT in severe patients, leading to increased adipose tissue activity. These findings explain the uncertainty in gray distribution and the asymmetry of value distribution in the EAT radiomics features.

*3.5.2 Based on uncertainty quantification*

Fig. 7 illustrates the uncertainty quantification results in the two validation cohorts. Compared to the hybrid model using lung radiomics features, the hybrid model incorporating both lung and EAT radiomics features exhibits a concentration of uncertainty levels in the ranges of 0-0.1 and 0.1-0.2, with higher accuracy within this range. This indicates that incorporating EAT radiomics features into the model enhances the confidence level and improves the accuracy of predicting the severity of patients. This may be attributed to the inclusion of EAT radiomic features likely captures additional characteristics related to the disease progression or underlying physiological factors that influence the severity of the condition. Additionally, it is noteworthy that the number of patients with uncertainty quantification levels between 0.5 and 1 is lower in the hybrid model using both lung and EAT radiomics features compared to the model using only lung radiomics features. Uncertainty levels exceeding 0.5 indicate that the hybrid model cannot provide more stable and confident predictions, necessitating secondary decision-making by physicians. This means that fewer secondary decision-making interventions by physicians were required for cases with high uncertainty levels. Therefore, incorporating EAT radiomic features not only enhanced the efficiency of clinical workflows but also suggested the potential for targeted interventions and personalized treatment strategies based on the model's predictions and additional insights from EAT radiomic features. Transparently presenting the uncertainty quantification results to physicians and patients can enhance their trust and acceptance of the model, making them more willing to embrace and adopt machine learning applications in COVID-19 severity classification.

## 4. Conclusions

This study proposed a three-stage EAT extraction method and demonstrated the incremental value and interpretability of EAT radiomics features for the severity of COVID-19 infection. The utilization of EAT radiomics features enhanced the classification accuracy and diagnostic effectiveness. Uncertainty quantification and radiomics features analysis confirmed the interpretability of case prediction after inclusion of EAT features.

**Acknowledgments**

This study received support from the National Natural Science Foundation of China (Grant Numbers: 62106233, 62303427, and 82370513), and the Henan Science and Technology Development Plan (Grant Number: 232102210010, 232102210062).

**Tables**

Table 1 Baseline Characteristics of Patients in the Derivation and Internal Validation Cohort Mild and Severe COVID-19.

| Characteristics | Derivation Cohort | | P | Internal Validation Cohort | | P |
|---|---|---|---|---|---|---|
| | Mild | Severe | | Mild | Severe | |
| Age, years, mean(±SD) | 39.98±15.39 | 59.23±14.86 | <.001 | 40.46±15.23 | 59.71±14.46 | <.001 |
| Gender, No. (%) | | | .59 | | | .15 |
|     Male | 125(48) | 16(53) | | 65(59) | 11(79) | |
|     Female | 135(52) | 14(47) | | 46(41) | 3(21) | |
| Past cardiovascular disease, No. (%) | 35(13) | 13(43) | <.001 | 16(14) | 5(36) | .045 |
| Lactate dehydrogenase, (U/L), mean(±SD) | 219.37±61.10 | 346.47±142.47 | <.001 | 220.29±83.26 | 326.29±102.63 | <.001 |
| NTPro-BNP, (pg/ml), mean(±SD) | 4.22±0.94 | 4.19±1.01 | .84 | 4.26±0.83 | 3.71±0.62 | .02 |
| Creatine kinase isoenzymes, (ng/ml), mean(±SD) | 13.19±12.07 | 16.43±10.69 | .16 | 12.61±4.66 | 12.91±2.37 | .82 |
| Hypertension, No. (%) | 31(12) | 9(30) | .01 | 16(14) | 5(36) | .045 |
| D-dimer, (ug/ml), mean(±SD) | 0.55±1.34 | 1.72±3.70 | <.001 | 0.57±1.64 | 2.23±4.96 | .01 |
| White blood cell, ($10^9$/L), mean(±SD) | 5.55±1.90 | 5.37±1.94 | .61 | 5.38±2.39 | 6.22±4.61 | .29 |
| Lymphocyte, ($10^9$/L), mean(±SD) | 1.44±0.60 | 1.02±0.41 | <.001 | 1.45±0.49 | 0.93±0.61 | <.001 |
| Serum sodium, (mmol/L), mean(±SD) | 139.90±2.70 | 137.70±3.77 | <.001 | 139.98±2.04 | 134.29±5.20 | <.001 |
| Urea, (mmol/L), mean(±SD) | 4.43±1.50 | 5.56±3.46 | <.001 | 4.43±1.13 | 5.04±2.09 | .09 |
| $PO_2$, (KPa), mean(±SD) | 14.34±4.34 | 9.94±4.94 | <.001 | 14.68±4.75 | 13.45±5.51 | .38 |
| $PCO_2$, (KPa), mean(±SD) | 5.43±0.45 | 5.49±0.89 | .66 | 5.39±0.59 | 4.94±0.72 | .01 |
| PCT, (ng/ml), mean(±SD) | 0.04±0.03 | 0.18±0.39 | <.001 | 0.04±0.04 | 0.16±0.18 | <.001 |
| APTT, (s), mean(±SD) | 38.57±6.35 | 42.06±6.57 | .01 | 38.02±4.10 | 43.10±5.69 | <.001 |
| PT, (s), mean(±SD) | 13.38±1.10 | 13.62±0.83 | .26 | 13.29±0.60 | 13.73±1.46 | .04 |
| Potassium, (mmol/L), mean(±SD) | 3.80±0.34 | 3.79±0.36 | .89 | 3.83±0.34 | 3.58±0.43 | .02 |
| Lactic acid, (mmol/L), mean(±SD) | 1.26±1.20 | 1.37±1.03 | .62 | 1.29±0.81 | 0.94±0.22 | .11 |
| HDL-C, (mmol/L), mean(±SD) | 29.23±3.90 | 28.46±3.77 | .31 | 29.02±3.80 | 26.26±3.14 | .01 |
| eGFR, (ml/(min*1.73$m^2$)), mean(±SD) | 117.13±25.09 | 106.31±29.54 | .03 | 118.49±22.74 | 108.55±29.74 | .14 |
| Creatinine, (umol/L), mean(±SD) | 65.30±19.22 | 69.14±33.25 | .35 | 64.35±14.09 | 70.31±21.95 | .17 |
| LDL-C, (mmol/L), mean(±SD) | 24.46±1.84 | 24.51±2.49 | .89 | 24.37±1.77 | 24.56±2.44 | .73 |
| ALT, (U/L), mean(±SD) | 25.03±18.11 | 29.23±19.50 | .24 | 29.72±22.25 | 36.08±35.10 | .36 |
| AST, (U/L), mean(±SD) | 24.61±19.05 | 37.50±25.24 | <.001 | 24.56±12.24 | 37.51±23.90 | <.001 |
| Dyspnea, No. (%) | 6(2) | 2(7) | .18 | 2(2) | 5(36) | <.001 |
| Diabetes, No. (%) | 14(5) | 3(10) | .31 | 4(4) | 3(21) | .01 |
| Antivirals, No. (%) | 161(56) | 29(97) | <.001 | 73(66) | 13(93) | .04 |
| Exudative lesions, No. (%) | 50(17) | 10(33) | .07 | 16(14) | 2(14) | .99 |
| CAD, No. (%) | 5(2) | 4(13) | <.001 | 0(0) | 0(0) | - |

ALT, alanine aminotransferase; APTT, activated partial thrombin time; AST, aspartate aminotransferase; CAD, coronary heart disease; eGFR, estimated glomerular filtration rate; HDL-C, high density liptein cholesterol; LDL-C, low density lipoprotein cholesterol; $PCO_2$, partial pressure of carbon dioxide; PCT, procalcitonin; $PO_2$, partial pressure of oxygen; NTPRO-BNP, NT pro B-type natriuretic peptide; PT, prothrombin time.

Table 2 Comparison of DSC and HD by different methods in two cohort test sets.

| | Method | DSC | HD(mm) |
|---|---|---|---|
| Cohort 1 | U-Net | 0.629(±0.047) | - |
| | V-Net | 0.921(±0.019) | 14.446(±3.143) |
| | Hoori et al.[18] | 0.935(±0.021) | 13.842(±3.486) |
| | Commandeur et al.[19] | 0.943(±0.016) | 10.548(±3.042) |
| | Yolo-V5 + U-Net (Ours) | **0.972(±0.011)** | **7.538(±2.112)** |
| Cohort 2 | U-Net | 0.653(±0.032) | - |
| | V-Net | 0.903(±0.024) | 17.169(±5.168) |
| | Hoori et al.[18] | 0.925(±0.022) | 14.846(±3.744) |
| | Commandeur et al.[19] | 0.937(±0.018) | 12.325(±4.894) |
| | Yolo-V5 + U-Net (Ours) | **0.968(±0.005)** | **6.423(±1.842)** |

DSC, Dice similarity coefficient; HD, Hausdorff distance.

Table 3 Features used in diagnostic models. Features were ranked according to the AUC in univariate analysis.

| Features of the source | Features |
|---|---|
| EAT | original_glszm_ZoneEntropy |
| Lung | original_firstorder_Kurtosis |
| Lung | original_glszm_SmallAreaEmphasis |
| EAT | original_firstorder_Skewness |
| Lung | original_glszm_LargeAreaLowGrayLevelEmphasis |
| EAT | original_glszm_LargeAreaLowGrayLevelEmphasis |
| Lung | original_glcm_MaximalCorrelationCoefficient |
| EAT | original_glszm_ZonePercentage |
| Lung | original_ngtdm_Strength |
| Lung | original_ngtdm_Complexity |

Table 4 Model Fitting and Calibration in the Derivation Cohort (n = 290) and the predictive performance of the internal validation cohort (n=125) and external validation cohort (n=90) for the mild and severe classification.

| Cohort | Features | SN | SP | AUC | ACC |
|---|---|---|---|---|---|
| Internal Validation | Lung | 0.786 | 0.910 | 0.867 | 0.896 |
| | Lung+EAT | **0.857** | **0.928** | **0.957** | **0.920** |
| External Validation | Lung | 0.742 | 0.847 | 0.851 | 0.810 |
| | Lung+EAT | **0.806** | **0.915** | **0.911** | **0.880** |

ACC, accuracy; AUC, area under the receiver operating characteristic curve; SN, sensitivity; SP, specificity.

Table 5 Model Performance Comparison with State-of-the-Art COVID-19 Severity Diagnosis Methods.

| Method | Features | Number of patients | AUC | ACC |
|---|---|---|---|---|
| Liang et al.[35] | Clinical | 1590 (131 severe) | 0.88 | - |
| Zhu et al.[5] | Clinical+ Lung radiomics | 427 (40 severe) | 0.94 | 0.93 |
| Zhao et al.[36] | Clinical | 172 (60 severe) | - | 0.91 |
| Zhang et al.[37] | Clinical | 422 (102 severe) | 0.90 | - |
| Li et al.[38] | CT visual quantitative | 78 (8 severe) | 0.92 | - |
| Our method | Lung + EAT radiomics | 415 (44 severe) | 0.96 | 0.92 |

ACC, accuracy; AUC, area under the receiver operating characteristic curve;

**Figures**

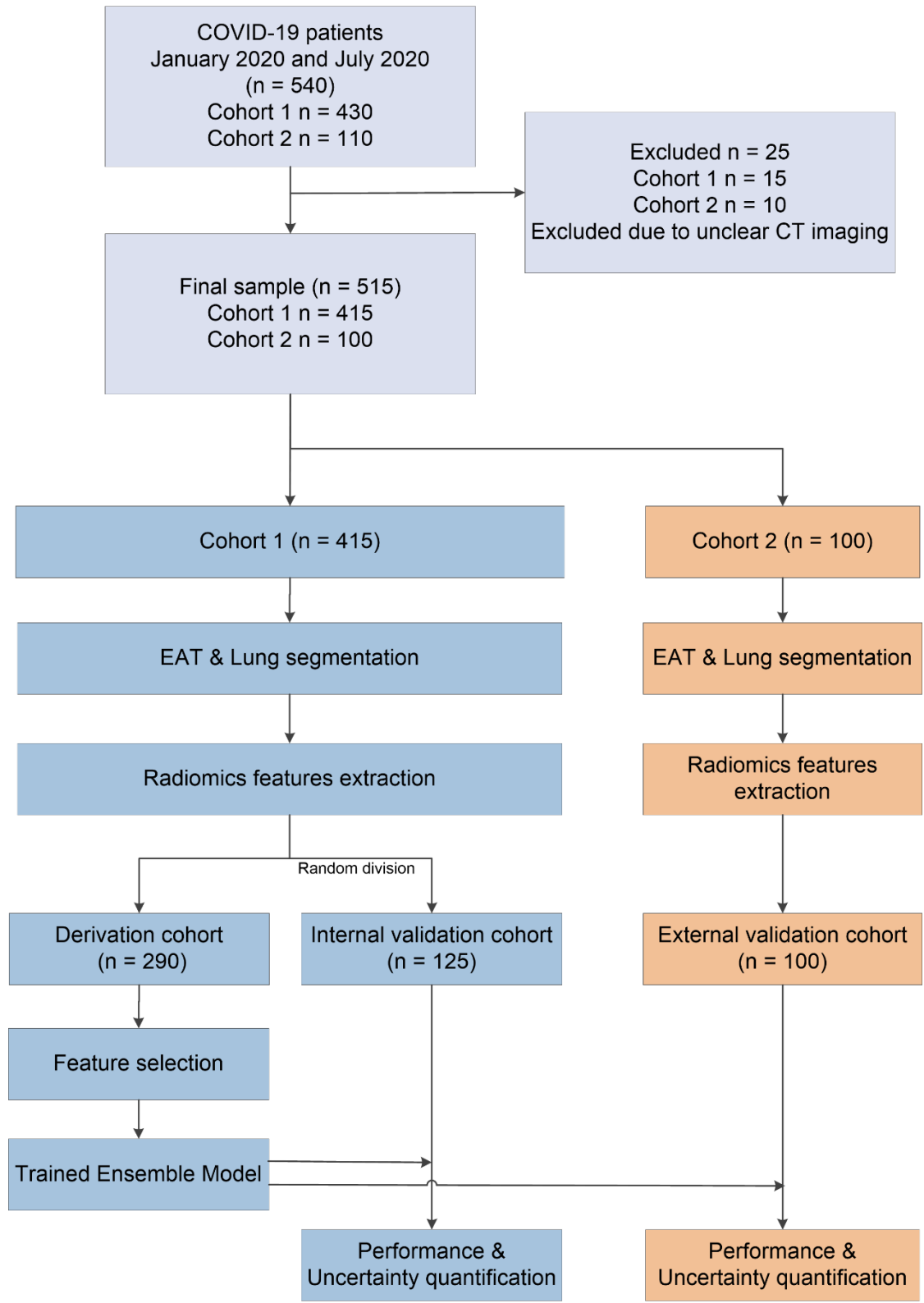

Fig. 1 Flowchart of patient inclusion and exclusion and flowchart of modeling.

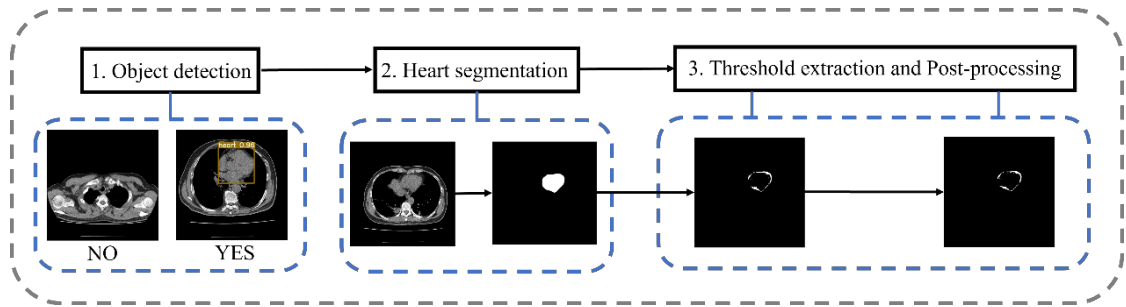

Fig. 2 Flowchart of EAT extraction.

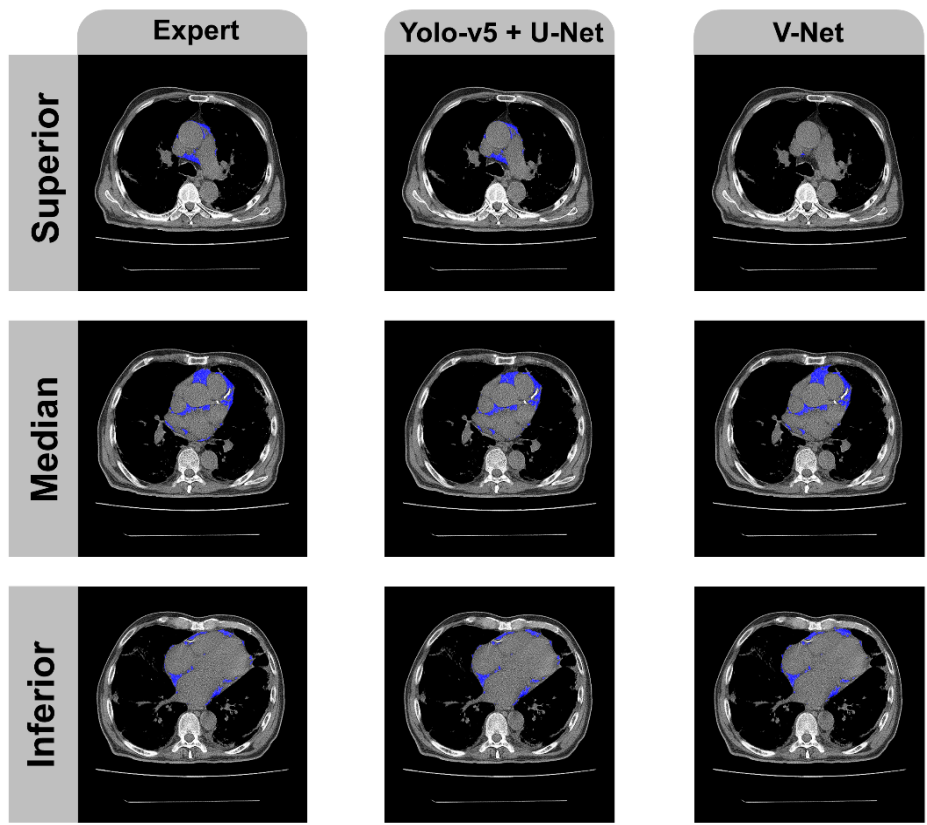

Fig. 3 Comparison of the results of EAT extraction.

Comparison of the results of EAT extraction based on expert segmentation (left column), automatic heart segmentation by Yolo-V5+U-Net method (middle column, DSC=0.972), and automatic heart segmentation by V-Net method (right column, DSC=0.921) in cohort 1. The blue area is EAT. The upper, middle, and lower rows correspond to the upper, middle, and lower parts of the heart.

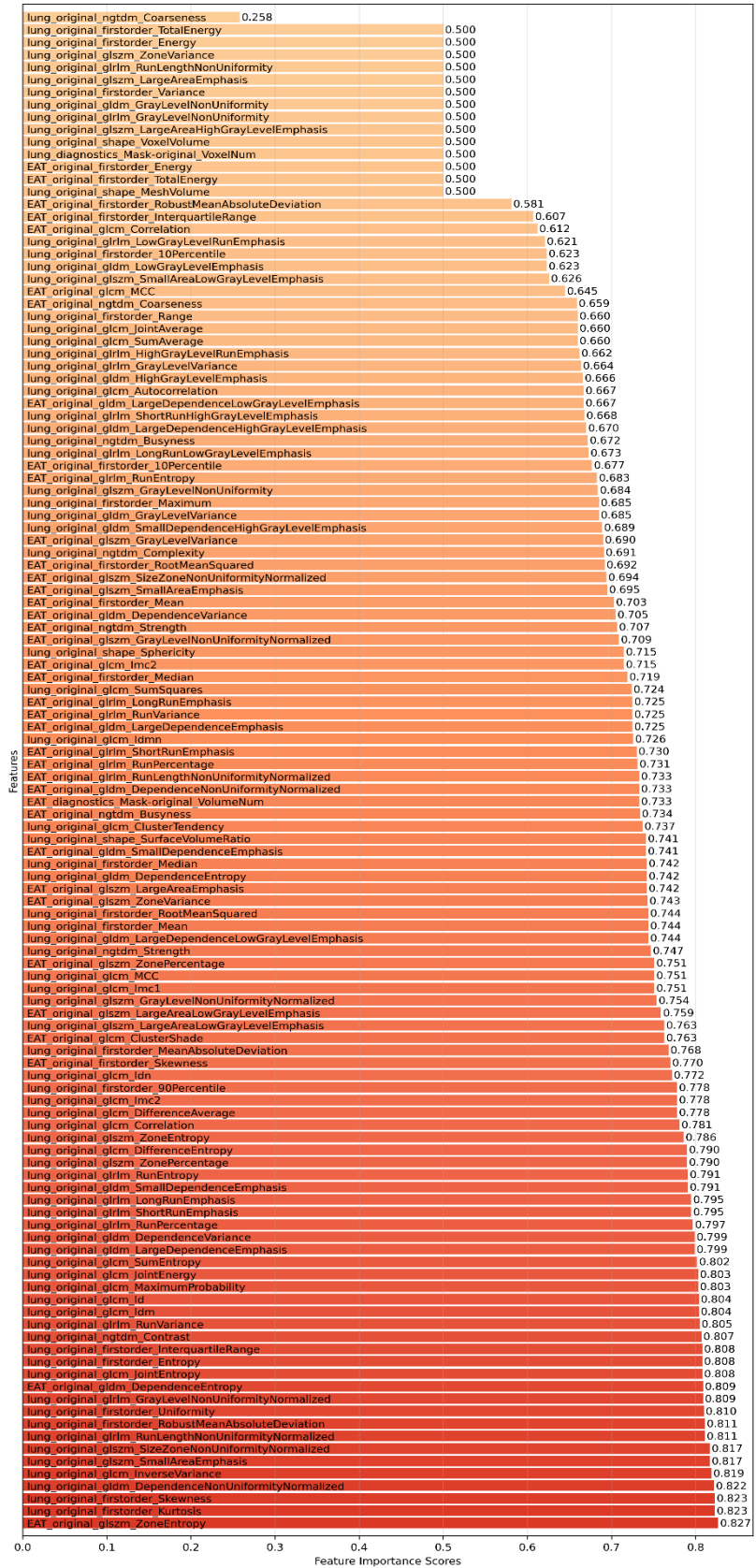

Fig. 4 Feature importance scores.

The left side of the bar shows the name of each feature, and the right side of the bar shows the feature importance score of the corresponding feature.

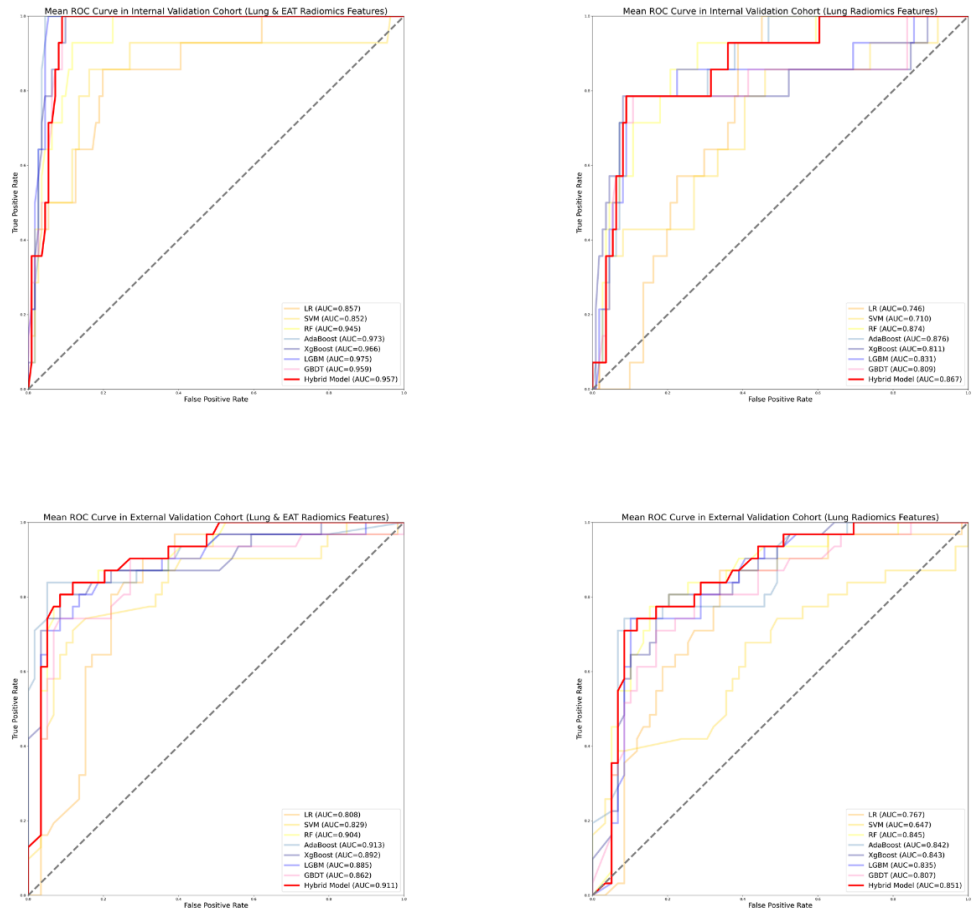

Fig. 5 AUC plots for Hybrid models with lung and EAT radiomics features in internal and external validation cohort.

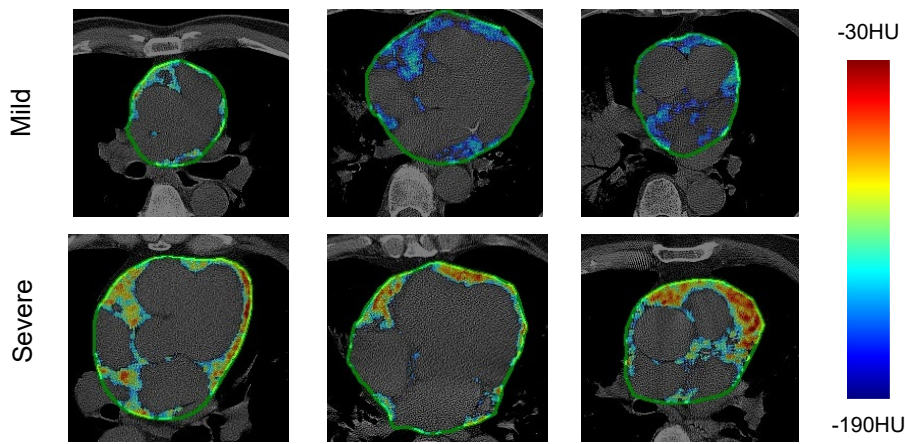

Fig. 6 Mapping of EAT attenuation heatmap for COVID-19 patients of different severity levels. The green contour represents the outline of the heart.

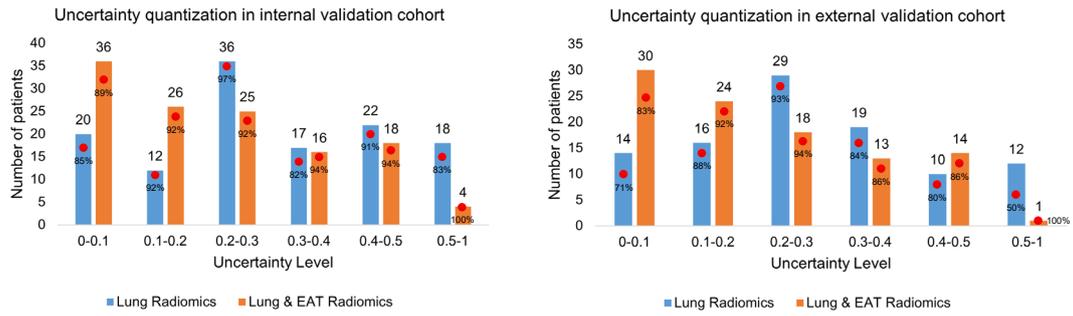

Fig. 7 Comparison of quantified levels of uncertainty between hybrid model of lung radiomics features and hybrid model of radiomics features of both lung and EAT in internal validation cohort and external validation cohort. The red dots are the prediction accuracy of each level.